\documentclass[12pt,preprint]{aastex}

\begin{document}

\title{Evidence of a Massive Black Hole Companion in the Massive Eclipsing Binary V Puppis}

\author{Qian S.-B.\altaffilmark{1,2,3}, Liao W.-P.\altaffilmark{1,2,3}, and
Fern\'{a}ndez Laj\'{u}s, E.\altaffilmark{4}}

\altaffiltext{1}{National Astronomical Observatories/Yunnan
Observatory, Chinese Academy of Sciences (CAS), P.O. Box 110, 650011
Kunming, P.R. China (e-mail: qsb@netease.com)}

\altaffiltext{2}{United Laboratory of Optical Astronomy, Chinese
Academy of Sciences (ULOAC), 100012 Beijing, P. R. China}

\altaffiltext{3}{Graduate School of the CAS, Beijing, P.R. China}

\altaffiltext{4}{Facultad de Ciencias Astron\'{o}micas y
Geof\'{i}sicas, Universidad Nacional de La Plata, Paseo del Bosque
s/n, 1900, La Plata, Pcia. Bs. As., Argentina}

\begin{abstract}
Up to now, most stellar-mass black holes were discovered in X-ray
emitting binaries, in which the black holes are formed through a
common-envelope evolution. Here we give evidence for the presence of
a massive black hole candidate as a tertiary companion in the
massive eclipsing binary V Puppis. We found that the orbital period
of this short-period binary (P=1.45 days) shows a periodic variation
while it undergoes a long-term increase. The cyclic period
oscillation can be interpreted by the light-travel time effect via
the presence of a third body with a mass no less than 10.4 solar
mass. However, no spectral lines of the third body were discovered
indicating that it is a massive black hole candidate. The black hole
candidate may correspond to the weak X-ray source close to V Puppis
discovered by Uhuru, Copernicus, and ROSAT satellites produced by
accreting materials from the massive binary via stellar wind. The
circumstellar matter with many heavy elements around this binary may
be formed by the supernova explosion of the progenitor of the
massive black hole. All of the observations suggest that a massive
black hole is orbiting the massive close binary V Puppis with a
period of 5.47 years. Meanwhile, we found the central close binary
is undergoing slow mass transfer from the secondary to the primary
star on a nuclear time scale of the secondary component, revealing
that the system has passed through a rapid mass-transfer stage.
\end{abstract}


\section{Introduction}

With an orbital period of 1.4545 days and two B-type components
stars (spectral types B1 and B3), V Puppis (HR\,3129 and HD 65818)
is one of a few massive semi-detached eclipsing binaries where the
less massive components are filling their Roche lobes (e.g.,
Andersen et al. (1983); Bell et al. (1987a, b); Terrell et al.
(2005)). According to the theory of binary evolution (e.g., Sybesma
1986), theses systems were formed through a case A mass-transfer
(mass transfer is occurring when hydrogen is burning in the core).
Since the discovery of V Puppis more than one century ago, it was
extensively investigated photometrically and spectroscopically (e.g.
Andersen et al. (1983) and references therein). A weak X-ray source
in the field of V Puppis (e.g., Giacconi et al. (1974); Groote et
al. 1978); Bahcall et al. (1975)) and a significant amount of
circumstellar matter around the binary (e.g., York et al. 1976 and
Koch et al. 1981) make it a very interesting binary to study.

Because of its stable light curve and the deep and sharp eclipsing
minima (e.g., Andersen et al. (1983)), times of light minimum of V
Puppis can be determined precisely. Thus, the variations of the
orbital period can in principle be derived with high precision by
analyzing the observed-calculated (O-C) diagram. In the present
letter, orbital period changes of V Puppis are investigated. Then,
based on the period variations, the presence of a massive black hole
companion and the evolutionary state of the system are discussed.

\section{Analysis of the orbital period changes}

Epoches and orbital periods of V Puppis have been given by several
authors (e.g., Andersen et al. 1983). Kreiner \& Ziolkowski (1978)
did not detect any period change for the binary, while Andersen et
al. (1983) pointed out that the period of V Puppis is variable.
However, since the early eclipse times were mainly observed
photographically or visually, conclusions on the period change are
not reliable. For the present analysis, we only use the
photoelectric (Pe) and charge-coupled device (CCD) observations.
Most of the data were collected by Kreiner et al. (2001). Two new
CCD times of light minimum were obtained on November 2, 2007 and on
March 24, 2008 with the Virpi S. Niemela 0.8-m telescope at La Plata
Observatory (see Table 1). During the observation, the V filter was
used. The $(O-C)_1$ values of all available Pe and CCD times of
light minimum were computed with the linear ephemeris given by
Kreiner et al. (2001),
\begin{equation}
Min.I=JD(Hel.)2428648.2837+1^{d}.45448686\times{E},
\end{equation}
\noindent where $2428648.2837$ is the time of conjunction,
$1^{d}.45448686$ the constant ephemeris period $P_{e}$, and $E$ the
cycle. The $(O-C)_1$ values are listed in the sixth column of Table
1 and are represented graphically against $E$ in the upper panel of
Figure 1.

If the orbital period of a binary varies linearly in time, the (O-C)
diagram should have a parabolic shape, as observed in other
eclipsing binaries (e.g., Qian \& Zhu 2006). As for V Puppis, a
simple second-order polynomial does not give a satisfactory fit (the
dashed line in the upper panel of Figure 1). Therefore, a sinusoidal
term was added to a quadratic ephemeris to give a good fit to the
$(O-C)_1$ curve. A least-squares solution leads to the following
ephemeris,
\begin{eqnarray}
Min.I&=& 2428648.2960(\pm0.0045)+
1^{d}.45448114(\pm0.00000091)\times{E}\nonumber\\
  & &+4.18(\pm0.44)\times{10^{-10}}\times{E^{2}}\nonumber\\
    & &+0.0163(\pm0.0016)\sin[0.^{\circ}2621(\pm0.0011)\times{E}+44.^{\circ}3(\pm3.^{\circ}9)].
\end{eqnarray}
\noindent The sinusoidal term suggests an oscillation with a period
of about $T=5.47$\,year and an amplitude of about $A=0.^{d}0163$,
which suggests a cyclic period oscillation with a period of 5.47
years (the upper panel of Fig. 1). Though the $(O-C)_1$ data do not
cover a whole cycle, as shown in Figure 1, the minimum, the maximum,
and the period of the cyclic variation are all well constrained,
indicating that the period changes are reliable. The quadratic term
in Eq. (2) reveals a continuous period increase at a rate of
$dP/dt=+2.10\times{10^{-7}}$\,days/year corresponding to a period
increase of 1.8\,s per century.

\section{Discussion of the orbital period variation}

\subsection{The presence of a massive black hole companion}

Both components of V Puppis are B-type stars containing a convective
core and a radiative envelope, which is different from the Sun.
Hence, the period oscillation can not be explained as solar-type
magnetic activity cycles (i.e. the Applegate mechanism) (Applegate
1992). Therefore, the simplest explanation of the cyclic period
oscillation is a wobble of the binary's barycentre due to the
presence of a tertiary companion. The $(O-C)_2$ values calculated
from the quadratic ephemeris in Eq. (2) are shown in the seventh
column of Table 1. The corresponding $(O-C)_2$ diagram is plotted
against the epoch number E in the middle panel of Figure 1. By
considering a general case with an elliptical orbit, the following
equation was used to describe those $(O-C)_{2}$ residuals,
\begin{equation}
(O-C)_{2}=a_{0}
+\sum_{i=1}^{2}{[a_{i}\cos(i{\Omega}{E})+b_{i}\sin(i{\Omega}{E})]}.
\end{equation}
\noindent The results were obtained with the same value of $\Omega$
($0.^{\circ}2621$) as that used in equation (2) (solid line in the
middle panel of Figure 1). The orbital parameters of the tertiary
companion were computed by using the formulae derived by Kopal
(1959),
\begin{equation}
a_{12}^{\prime}\sin{i}^{\prime}={\it c}\sqrt{a_{1}^{2}+b_{1}^{2}},
\end{equation}

\begin{equation}
e^{\prime}=2\sqrt{\frac{a_{2}^{2}+b_{2}^{2}}{a_{1}^{2}+b_{1}^{2}}},
\end{equation}

\begin{equation}
{\omega}^{\prime}=\arctan\frac{(b_{1}^{2}-a_{1}^{2})b_{2}+2a_{1}a_{2}b_{1}}
                              {(a_{1}^{2}-b_{1}^{2})a_{2}+2a_{1}b_{1}b_{2}},
\end{equation}

\begin{equation}
{\tau}^{\prime}=t_{0}-\frac{T}{2\pi}
                \arctan\frac{a_{1}b_{2}-b_{1}a_{2}}{a_{1}a_{2}+b_{1}b_{2}},
\end{equation}

\noindent where {\it c} is the speed of light, $a_{12}^{\prime}$,
$i^{\prime}$, $e^{\prime}$, ${\omega}^{\prime}$, and
${\tau}^{\prime}$ are the semi-major axis, the orbital inclination,
the eccentricity, the longitude of the periastron from the ascending
node, and the time of the periastron passage, respectively. The
solutions are displayed in Table 2. With the following well-known
equations,
\begin{equation}
f(m) =\frac{4\pi^{2}}{G{T}^{2}}\times(a_{12}\sin{i^{\prime}})^{3}
\end{equation}
and
\begin{equation}
f(m)=\frac{(M_{3}\sin{i^{\prime}})^{3}} {(M_{1}+M_{2}+M_{3})^{2}},
\end{equation}
\noindent the mass function, the masses and radii of the third body
were calculated using the absolute parameters determined by Andersen
et al. (1983). In the formula, $M_{1}$, $M_{2}$, and $M_{3}$ are the
masses of the eclipsing pair and the third body, respectively, G is
the gravitational constant, and T is the period of the (O-C)
oscillation. For different values of the orbital inclination, the
masses and orbital radii of the tertiary component are displayed in
Table 2. The corresponding relations between the mass and the radius
of the third body and the orbital inclination are shown in Figure 2.

The lowest mass of the tertiary companion is $10.4$\,$M_{\odot}$,
i.e. larger than the mass of the secondary component. If the
tertiary component is a normal star, we would see its spectral lines
not changing with the orbital phase of the binary. However, this has
not been reported (e.g., York et al. 1976; Koch et al. 1981;
Andersen et al. 1983). Moreover, the third body is more massive than
the secondary component ($M_2=7.76$\,$M_{\odot}$) in the central
binary, it should be very luminous and should contribute a large
amount of third light to the total system unless it is not a normal
main-sequence star. Therefore, we suspect that the tertiary
component is a compact object and a probable black hole candidate.
This situation resembles that in the triple system HR 2876 where a
compact object (possibly a white dwarf star) with a mass excess of
1.0\,$M_{\odot}$ was found to be orbiting a B-type close binary star
(B3.5V+B6V; P=15 days) (Burleigh \& Barstow 1998; Vennes 2000). The
main differences between the two systems are that the tertiary
companion in the V Puppis is much more massive and that the central
eclipsing binary in V Puppis system has a much more tight orbit and
shows strong interaction between both components.

A weak X-ray source (3U 0750-49) with $9.4\pm2.3$ count $s^{-1}$ was
detected by the Uhuru satellite (Giacconi et al. 1974). The
counterpart of the X-ray source was a puzzling problem for
astronomers since the 70s of last century. V Puppis was initially
listed in the 3U catalogue as a possible candidate by Giacconi et
al. (1974). However, another bright star HD\,64740, close to the
source (within the 90 percent confidence error box) was also
considered as a possible candidate (Groote et al. 1978). By
analyzing the X-ray data of V Puppis and HD 64740 from the
Copernicus satellite, Bahcall et al. (1975) later concluded that the
weak X-ray source must almost certainly be closer to V Puppis than
to HD 64740, but most probably neither V Puppis nor HD 64740 is the
candidate. More recently, V Puppis was also discovered to be a X-ray
source by ROSAT satellite with $0.08$ count $s^{-1}$ (X-ray
50\,{\AA}, 250\,eV) (e.g., Lampton et al. 1996). Our study indicates
that the tertiary companion may correspond to the weak X-ray source.
There are two mechanisms that can produce the X-ray emission in the
V Puppis system. If the massive tertiary companion is a normal star,
the X-ray emission could be caused by the colliding winds from the
eclipsing pair and the tertiary companion, resembling those observed
in other massive binary systems (e.g., Stevens 1992; Pittard \&
Stevens 2002). If the third body is a massive black hole, the X-ray
source is reproduced by accreting material from the massive binary
via stellar wind as observed in X-ray emitting binaries (e.g.,
Prestwich et al. 2007). However, the former mechanism can be ruled
out, because no spectral lines of the tertiary component were found.

Another interesting feature of V Puppis is the existence of a
significant amount of circumstellar matter, which was deduced from
the optical spectra and ultraviolet observations obtained by the IUE
spacecraft and the Copernicus satellite (e.g., York et al. 1976;
Koch et al. 1981; Andersen et al. 1983). Numerous interstellar
spectral lines, especially those of many types of heavy elements
(e.g., Mg II, Si I, Si II, Si III, C I, O I, Fe II and Fe III) were
found in the H II region around V Puppis, in which strong lines of
Si III and Fe III are remarkable (e.g., York et al. 1976 and Koch et
al. 1981). The formation of the circumstellar matter around V Puppis
is an unsolved problem. Considering that V Puppis has undergone mass
exchange, one may think that the circumstellar matter results from
from the binary evolution. However, we may question why the other
systems (e.g. AI Cru) that resemble V Puppis do not have the same
significant amount of circumstellar matter (Bell et al. 1987). The
most probable reason for the formation of the circumstellar matter
around V Puppis is that it is formed by the supernova explosion of
the progenitor of the tertiary black hole companion (e.g., Woosley
\& Weaver (1995) and Zhang et al. (2007)). This may produce the
heavy elements in the circumstellar matter and results in the
formation of the H II region around the eclipsing binary.

\subsection{The slow mass transfer of V Puppis}

V Puppis is a semi-detached binary system with the secondary
component filling the critical Roche lobe (e.g., Andersen et al.
1983). The long-term period increase can be explained by the mass
transfer from the secondary to the primary component. By using the
equation,
\begin{equation}
{\tau}_{N}=10^{10}M_2/L_2,
\end{equation}
\noindent where $M_2$ and $L_2$ are the masses and the luminosity of
the less massive component, the nuclear time-scale of the secondary
is calculated to be ${\tau}_{N}=7.1\times10^{6}$\,years. The value
is close to the timescale of the period change
${\tau}_{P}=\frac{P}{\dot{P}}=6.9\times10^{6}$\,years, which reveals
that V Puppis is now undergoing a slow mass-transfer evolutionary
stage on the nuclear time-scale of the less massive component. This
suggests that V Puppis is formed via a case A evolution and has
passed through a rapid mass-transfer stage. The results are in
agreement with the prediction from the theory of massive binary
evolution (e.g., Sybesma 1986). By assuming a conservative mass
transfer, the well-known equation
\begin{equation}
\frac{\dot{P}}{P}=3\dot{M}(\frac{1}{M_{2}}-\frac{1}{M_1}),
\end{equation}
\noindent yields a rate of mass transfer
$dM/dt=7.82\times{10^{-7}}$\,$M_{\odot}/year$. However, since the
nuclear time-scale of the secondary is slightly longer than the time
scale of period change, the slow mass transfer is insufficient to
cause the observed period increase. This suggests that the stellar
wind from the massive component of the binary should contribute to
the period increase. The situation resembles that of AI Cru (Qian et
al. 2008).

By considering conservative mass and angular momentum, the
calculation by Plavec (1968) suggests that, to produce a case A mass
transfer of massive binaries, the initial orbital period should less
than 1.8\,days. One may ask a question what is the reason that
causes the origin of the initially short-period detached system? It
is possible that the massive tertiary companion has played an
important role for the origin and evolution of inner binary by
removing angular momentum from the central system via Kozai
oscillation (Kozai 1962) or a combination of Kozai cycle and tidal
friction (e.g., Fabrycky \& Tremaine 2007). This makes the inner
eclipsing pair to have a lower angular momentum and a shorter
initial orbital period.

\section{Conclusions}

The cyclic change of the orbital period, the presence of a weak
X-ray source near V Puppis, and the significant amount of
circumstellar matter around the binary star all support the
conclusion that a massive black hole is orbiting the massive close
binary V Puppis with a period of 5.47 years. Several massive stellar
black hole candidates have been discovered recently (e.g., Greiner
et al. (2001) and Orosz (2007)), but all candidates are found to be
a component in a binary system. We conclude here that we have
provided observational evidence for the first massive stellar black
hole candidate as a tertiary companion of a massive close binary
star. Theoretical investigations have shown that the formation of
black hole binaries should experience a common-envelope phase with
the spiral-in of the companion in the envelope causing the ejection
of the envelope (e.g., Taam \& Sandquist (2000) and Podsiadlowski et
al. (2002)). The mean distance between the black hole candidate and
the eclipsing pair is about 5.5 astronomical units (AU). During the
supergiant phase, the radius of the precursor of the black hole
companion in V Puppis may reach to several AUs. The survival of V
Puppis in this evolutionary stage suggests that the tertiary black
hole companion must be formed through an evolutionary process that
is different from that of the black holes in X-ray binaries (e.g.,
Woosley \& Weaver (1995) and Zhang et al. (2007)) and it could give
some constraints to the evolution of single massive stars and the
formation of isolated black holes.

From the present investigation, we deduce that the progenitor of V
Puppis is a triple system that is composed of a short-period
detached B-type eclipsing binary and a much more massive tertiary
companion. By drawing angular momentum from the inner binary through
Kozai oscillation (Kozai 1962) or a combination of Kozai cycle and
tidal friction (e.g., Fabrycky \& Tremaine 2007), the massive
tertiary companion may play an important role for the origin the
central system, which causes the inner eclipsing binary to have a
very short initial orbital period. In the triple system, the massive
tertiary companion evolves faster, and finally into a massive black
hole via a supernova explosion (e.g., Woosley \& Weaver (1995) and
Zhang et al. (2008)). At the same time, the original more massive
component of the eclipsing pair evolves to fill its critical Roche
lobe. Then, the binary was undergoing a rapid mass-transfer stage,
and after the mass ratio was revered, the central binary finally
reached the present semi-detached configuration with a slow
secondary-to-primary mass transfer on the nuclear time scale of the
present less massive star. Supergiant stars have recently been found
to be tertiary companions of OB-type close binaries in the two
triple systems HD 167971 and FR Sct (e.g., Leitherer et al. (1987);
Davidge \& Forbes (1988), and Pigulski \& Michalska (2007)). They
resemble the progenitor of V Puppis with the tertiary supergiant
companion evolving into a compact object. As for V Puppis, the
existence of the black hole in the outer orbit poses strong
constrain on the formation of the black hole, the mass loss in the
supernova that lead to the formation of the black hole, and the
magnitude of the kick received by the black hole, all make the
object a much more interesting system for future study.

\acknowledgments{This work is partly supported by the Yunnan Natural
Science Foundation (No. 2005A0059M), Chinese Natural Science
Foundation (No.10573032 and No.10433030), and the National Key
Fundamental Research Project through grant 2007CB815406. New CCD
photometric observations of V Puppis were obtained with the Virpi S.
Niemela 0.8-m telescope at La Plata Observatory. We are grateful to
Prof. Kreiner for sending us the times of light minimum of the
binary and to Cecilia Fari\~{n}a and Carolina von Essen for their
helps during the observations and the data processing. The authors
thank Prof. J.-P. De Greve for improving the English writing of the
original manuscript and the referee for those useful comments and
suggestions that help to improve the paper.}

\clearpage

\begin{table}
\caption{All available photoelectric and CCD times of light minimum
for V Puppis.}
\begin{tabular}{lllllllll}\tableline
JD.Hel. & Errors & Meth. & Min. & E & $(O-C)_1$ & $(O-C)_2$ &
Residuals & Ref. \\
2400000+ & days & & &  & days & days & days &\\\tableline
31917.957 & $\pm0.0010$ & Pe & I & 2248 & -0.0132 & -0.0147 & -0.0004 & (1)\\
31965.952 & $\pm0.0010$ & Pe & I & 2281 & -0.0162 & -0.0176 & -0.0046 & (1)\\
39869.6264& $\pm0.0010$ & Pe & I & 7715 & -0.0234 & -0.0164 & -0.0011 & (1)\\
41791.0113& $\pm0.0002$ & Pe & I & 9036 & -0.0157 & -0.0104 & +0.0065 & (2)\\
42537.9043& $\pm0.0001$ & Pe & II&9549.5& -0.0017 & +0.0026 & -0.0042 & (2)\\
42561.9044& $\pm0.0005$ & Pe & I &9566  & -0.0006 & +0.0036 & -0.0044 & (2)\\
43124.8032& $\pm0.0002$ & Pe & I &9953  & +0.0118 & +0.0151 & -0.0001 & (2)\\
43148.8024& $\pm0.0002$ & Pe & II&9969.5& +0.0119 & +0.0151 & +0.0014 & (2)\\
43151.7111& $\pm0.0002$ & Pe & II&9971.5& +0.0117 & +0.0149 & +0.0014 & (2)\\
45367.6063& $\pm0.0002$ & Pe & I &11495 & -0.0039 & -0.0056 & -0.0022 & (2)\\
45383.6052& $\pm0.0001$ & Pe & I &11506 & -0.0043 & -0.0061 & -0.0014 & (2)\\
48391.5025& $\pm0.0010$ & CCD& I &13574 & +0.0142 & +0.0026 & +0.0021 & (3)\\
48425.6868& $\pm0.0007$ & CCD& II&13597.5&+0.0180 & +0.0063 & +0.0044 & (3)\\
48500.5947& $\pm0.0050$ & Pe & I &13649  &+0.0198 & +0.0078 & +0.0025 & (4)\\
49821.2500& $\pm0.0050$ & Pe & I &14557  &+0.0011 & -0.0164 & -0.0003 & (4)\\
54456.7420& $\pm0.0010$ & CCD& I &17744  &+0.0435 & +0.0012 & -0.0024 & (5)\\
54550.5667& $\pm0.0010$ & CCD& II&17808.5&+0.0538 & +0.0109 & +0.0028 & (5)\\
\tableline
\end{tabular}\\
{References in Table 1:\\
(1) Kreiner \& Ziolkowski (1978); (2) Andersen et al. (1983); (3)
Kreiner (2006); (4) Stickland (1998); (5)The present authors.}
\end{table}

\begin{table}
\caption{Calculated orbital parameters of the tertiary companion in
V Puppis.}
\begin{tabular}{lll}\tableline\tableline
 Parameters    & Values & Errors \\\tableline
$a_{0}$        & -0.0005& $\pm0.0001$\\
$a_{1}$        & +0.0142& $\pm0.0012$\\
$b_{1}$        & +0.0113& $\pm0.0012$\\
$a_{2}$        & -0.0031& $\pm0.0019$\\
$b_{2}$        & +0.0026& $\pm0.0004$\\\tableline
$a_{12}^{\prime}\sin{i}^{\prime}$ (AU) & 3.14 & $\pm0.21$\\
$e^{\prime}$                         & 0.45 & $\pm0.16$\\
${\omega}^{\prime}$ & $-63.^{\circ}0$ & $\pm18.^{\circ}9$\\
${\tau}^{\prime}$ (days) & 2428213   & $\pm101$\\
f(m) ($M_{\odot}$)             & $1.03$   & $\pm0.21$\\
$M_{3}(i^{\prime}=90^{\circ})$ & $10.40$ & $\pm0.87$\\
$M_{3}(i^{\prime}=70^{\circ})$ & $11.25$ & $\pm0.96$\\
$M_{3}(i^{\prime}=50^{\circ})$ & $14.74$ & $\pm1.33$\\
$M_{3}(i^{\prime}=30^{\circ})$ & $27.44$ & $\pm2.87$\\
$d_{3}(i^{\prime}=90^{\circ})$ & $6.82$ & $\pm0.73$\\
$d_{3}(i^{\prime}=70^{\circ})$ & $6.80$ & $\pm0.73$\\
$d_{3}(i^{\prime}=50^{\circ})$ & $6.28$ & $\pm0.70$\\
$d_{3}(i^{\prime}=30^{\circ})$ & $5.56$ & $\pm0.66$\\
\tableline\tableline
\end{tabular}
\end{table}

\clearpage

\begin{figure}
\begin{center}
\includegraphics[angle=0,scale=1.2]{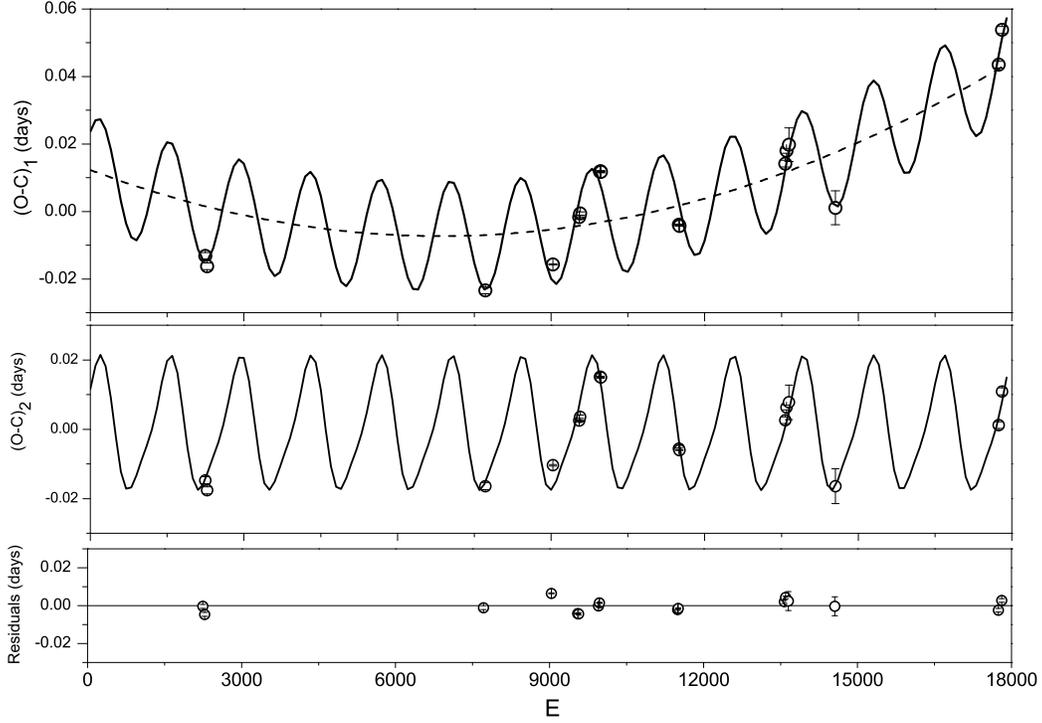}
\caption{A plot of the $(O-C)_1$ curve of V Puppis with respect to
the linear ephemeris given by Kreiner et al. (2001). The solid line
in the upper panel suggests a combination of a long-term period
increase and a cyclic change, while the dashed line refers to the
continuous increase of the orbital period. The middle panel shows
the plot of the $(O-C)_2$ residuals of the eclipse timings based on
the quadratic ephemeris in Eq. (1). The solid line in the middle
panel is the theoretical orbit of the massive black hole candidate
with an eccentricity of 0.45. Residuals after subtracting all
effects of period change are displayed in the lower panel, and no
variations can be traced there.}
\end{center}
\end{figure}

\begin{figure}
\begin{center}
\includegraphics[angle=0,scale=1.2]{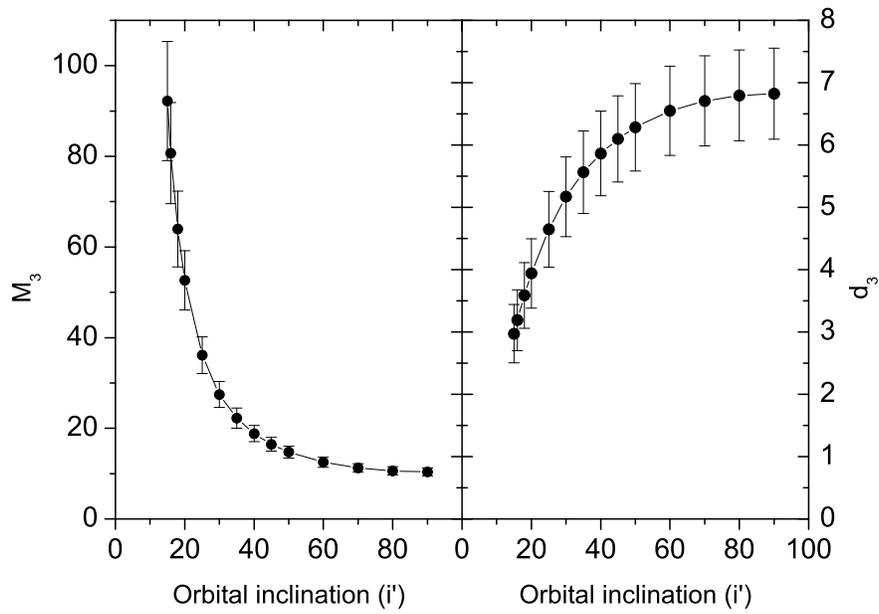}
\caption{The relations between the mass $M_3$\,($M_{\odot}$) and the
orbital radius $d_3$ (AU) of the tertiary component and its orbital
inclination $i^{\prime}$ in the V Puppis system. The tertiary
companion should be more massive than the secondary component and
thus may be a black hole candidate.}
\end{center}
\end{figure}

\end{document}